\documentclass[pra,aps,showpacs,
twocolumn]{revtex4}
\usepackage{epsfig}
\usepackage{epsfig}
\usepackage{graphicx}
\usepackage{amssymb}
\usepackage[usenames,dvipsnames]{color}

\begin{document} 
\title{Low-threshold Optical Parametric Oscillations in a Whispering Gallery Mode Resonator}

\author
{J. U. F{\"u}rst,$^{1,2}$ D. V. Strekalov,$^{1,3}$ D. Elser,$^{1,2}$ A. Aiello,$^{1,2}$ U. L. Andersen,$^{1,4}$ \\Ch. Marquardt,$^{1,2}$ and G. Leuchs$^{1,2}$\\
\normalsize{$^1$Max Planck Institute for the Science of Light, Erlangen, Germany,}\\ 
\normalsize{$^2$Department of Physics, University of Erlangen-Nuremberg, Germany,}\\
\normalsize{$^3$Jet Propulsion Laboratory, California Institute of Technology, Pasadena, USA,}\\
\normalsize{$^4$Department of Physics, Technical University of Denmark, Kongens Lyngby, Denmark.}
\\
}

\date{\today}


\begin{abstract}

In whispering gallery mode (WGM) resonators light is guided by continuous total internal reflection along a curved surface. Fabricating such resonators from an optically nonlinear material one
takes advantage of their exceptionally high quality factors and small mode volumes to achieve
extremely efficient optical frequency conversion. Our analysis of the phase matching conditions for optical parametric down conversion (PDC) in a spherical WGM resonator shows their direct relation to the sum rules for photons' angular momenta and predicts a very low parametric oscillations threshold. We realized such an optical parametric oscillator (OPO) based on naturally phase-matched PDC in Lithium Niobate. We demonstrated a single-mode, strongly non-degenerate OPO with a threshold of 6.7 $\mu$W and linewidth under 10 MHz. 
This work demonstrates the remarkable capabilities of WGM-based OPOs and opens the perspectives for their applications in quantum and nonlinear optics, particularly for the generation of squeezed light. 
\end{abstract}


\maketitle 

WGM resonators have found applications in many areas of photonics, including quantum and nonlinear optics, spectroscopy, biophysics and optomechanics  (see \cite{vahala1,ilchenko_app,optomech} and references therein). These areas benefit from a high quality factor ($Q$) and strong field confinement of WGM resonators in two different aspects. First, WGM resonators provide strong coupling of the confined field to other systems such as quantum dots \cite{gotzinger}, atoms \cite{aoki}, mechanical oscillators \cite{vahala_rev} or other cavities \cite{grudinin}. Second, WGM resonators provide strong nonlinear coupling between optical fields. Here the possibilities lie in utilizing either third order nonlinear effects in fused silica \cite{kippenberg_4wave,carmon05_locking,delhaye_4wave}, or nonlinear effects in crystalline materials \cite{savchenkov_4wave,savchenkov_07,ilchenko04SH,furst10}.   

In this letter we demonstrate optical parametric oscillations with extremely low thresholds and narrow linewidths in a crystalline WGM disk resonator. This is achieved by taking advantage of the strong second-order optical nonlinearity of Lithium Niobate and of the high quality factor of our WGM resonator, limited by the material absorption only. This research is motivated by the perspective to develop a miniature, stable, narrow-line tunable OPO for spectroscopy applications, to study the OPO dynamics far above the threshold, and by the
wide and successful use of conventional OPOs for the generation of non-classical light.  

To achieve efficient parametric processes in a nonlinear material phase matching conditions have to be fulfilled. For parametrically interacting plane waves in a bulk material, these phase matching conditions correspond to conservation of energy and \textit{linear} momentum of pump (p), signal (s) and idler (i) photons:
\begin{equation}
\hbar\omega_p = \hbar\omega_s+\hbar\omega_i,\quad \hbar{\vec k}_p = \hbar{\vec k}_s+\hbar{\vec k}_i.
\label{conserv}
\end{equation} 
In spherical geometry the \textit{orbital angular} momentum of the photons is conserved, along with the energy. This gives rise to WGM selection rules for the PDC process and determines the OPO threshold in-coupled pump power for various modes \cite{ilchenko03parametric}: 
\begin{equation}
P_{th}=
\frac{c}{\lambda_pQ_p}\left(\frac{n^3}{16\pi\sigma \chi^{(2)}Q_s}\right)^2,
\label{threshold}
\end{equation} 
where $c$ is the speed of light, $\lambda_p$ is the pump wavelength $n = n_e(\lambda_p)=n_o(\lambda_s)$ is the phase-matched refractive index, and $\chi^{(2)}$ is the effective quadratic nonlinearity. 
Let us point out that the OPO threshold power (\ref{threshold}) is closely related to the second harmonic saturation power $P_0 = 4 P_{th}$ \cite{ilchenko04SH}, provided that the  same WGMs are used.

The factor $\sigma$ in (\ref{threshold})
is determined by the overlap of the pump, signal and idler WGM normalized eigenfunctions $\Psi_{p,s,i}$. In spherical geometry these functions are expressed via spherical Bessel functions $j_\ell(k_qr)$ and spherical harmonics $Y_{lm}(\theta,\varphi)$ and are characterized by the azimuthal, polar and radial mode numbers $m$, $\ell$ and $q$. For large $m$ and $\ell$ the overlap factor in Eq. (\ref{threshold}) breaks up into a product of the radial part  $\sigma_r$ and the angular part expressed in terms of the Clebsch-Gordan coefficients \cite{Kozyreff08}:
\begin{equation}
\sigma =\int \Psi_s\Psi_i\Psi^*_p\,dV = \sigma_r\cdot \langle \ell_s ,\ell_i; m_s ,m_i| \ell_{p},m_{p}\rangle.
\label{overlap}
\end{equation} 
The radial overlap factor $\sigma_r$
is found numerically, approximating the large-index Bessel functions in vicinity of their $q$-th zero by Airy functions.
It turns out that for large values of $\ell_{s,i,p}$ and $m_{s,i,p}$ accessible in our experiment, $\sigma_r$ is nearly constant. Its dependence on $q_{s,i}$ is shown in Fig.~\ref{fig:radial} for 
$q_p = 1$. The maximum of $\sigma_r \approx 230\,{\rm cm}^{-3/2}$ is reached for $q_s=q_i= 1$, however for other $q_{s,i}$ the overlap remains significant. 

\begin{widetext}
\begin{figure}[h]
	\centering \hspace*{1cm}
	\includegraphics[width=16cm]{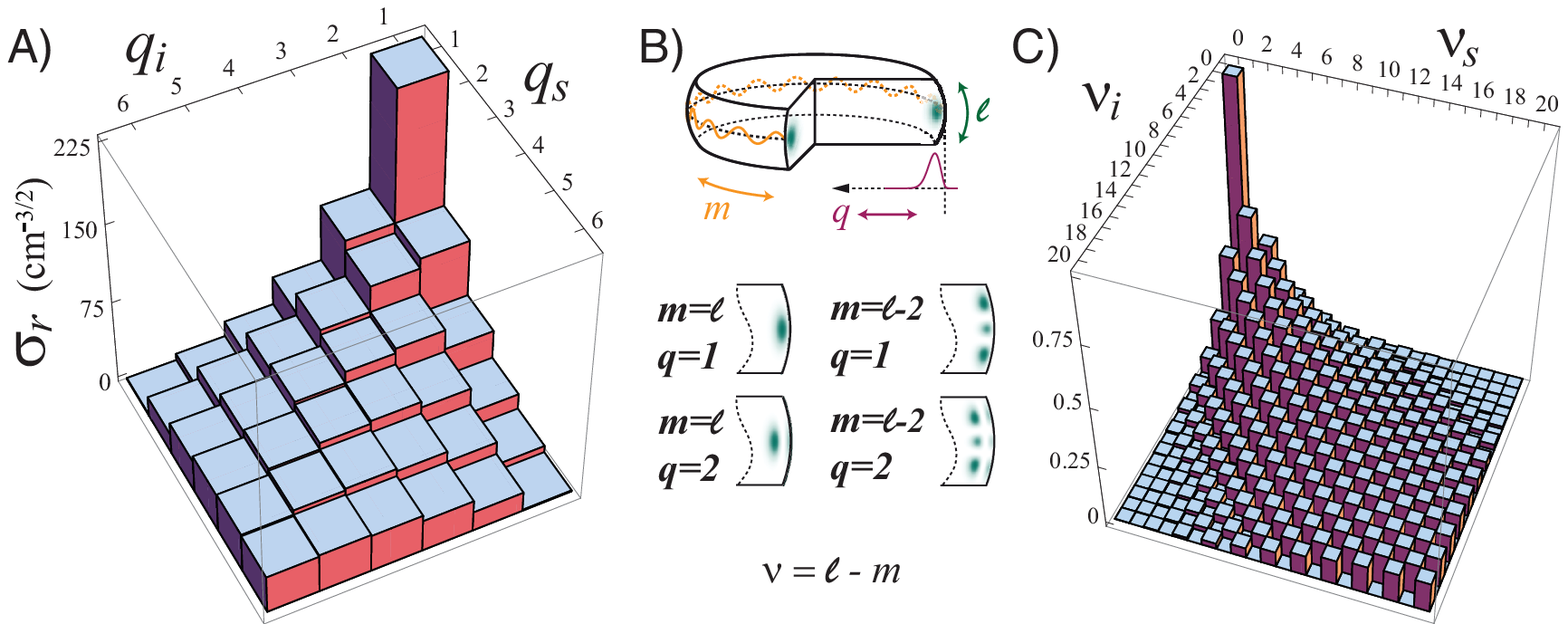}
\caption{\mbox{(Color online.) A) The radial overlap factor for the pump WGM nearest to the surface ($q_p = 1$) and various signal}
\mbox{and idler WGMs. B) The intensity distribution inside a WGM resonator for various mode numbers $\ell,\; m\; q$. C) The norma-}
\mbox{lized angular overlap factor for an equatorial pump WGM and various signal and idler WGMs.}}
	\label{fig:radial}
\end{figure} \vspace*{-0.1in}
\end{widetext}

The Clebsch-Gordan coefficients arise when orbital momenta are added, e.g. in atomic physics. Their presence in (\ref{overlap}) indicates that in the PDC process in spherical geometry the pump photon's orbital momentum is shared by the signal and idler photons. The selection rules for this process replace the free-space phase matching conditions given by the second equation in (\ref{conserv}):
\begin{eqnarray}
&m_{s}+m_i=m_p&\nonumber\\
& \quad |\ell_{s}-\ell_i|\leq \ell_p\leq \ell_s+\ell_i&\label{m} \\
&\ell_p+\ell_s+\ell_i = 0, \, 2, \, 4\,...&\nonumber
\end{eqnarray}

To simplify the analysis of the orbital selection rules, we assume that in addition to being closest to the surface, which means that $q_p = 1$, the pump WGM is also equatorial, i.e. $\ell_p = m_p$. We evaluate the Clebsch-Gordan coefficients using Gaunt's formula \cite{Gaunt29}, treating large factorials in Stirling's approximation. The coefficients coupling the equatorial signal and idler WGMs ($\ell_s = m_s$, $ \ell_i = m_i$) to the equatorial pump WGM ($\ell_p = m_p = m_s+m_i$) only weakly depend on $m_p$ and are practically independent of how it is shared between the signal and idler photons: 
\begin{equation}
\langle m_s, m_i ;m_s, m_i | m_p,m_p\rangle\approx \frac{1}{2} \left(m_p/\pi^{3}\right)^{1/4}. \label{norm}
\end{equation}
In other words, the normalized angular overlap factor
$
\langle m_s+\nu_s, m_i+\nu_i ;m_s, m_i | m_p,m_p\rangle\cdot 2\left(\pi^{3}/m_p\right)^{1/4} 
$
depends only on $\nu_s = \ell_s-m_s$ and $\nu_i=\ell_i-m_i$, as shown in Fig.~\ref{fig:radial}. Again, while the all-equatorial (i.e., $\nu_s=\nu_i=0$) WGMs coupling is the strongest, coupling of other signal and idler WGMs to the equatorial pump is also significant.
Therefore we expect, and observe in experiment, several PDC channels for the same pump mode. Different $\sigma$ and $Q$-factors of these modes lead to a different OPO threshold for each channel. For the most efficient all-equatorial case 
our analysis based on the measured $Q_{s,i}\approx 3.4\times 10^7$ and $Q_{p}\approx 2.0\times 10^7$ yields threshold power $P_{th}\approx 13$ nW. 

\begin{figure}[b]
	\centering
	\includegraphics[width=7cm]{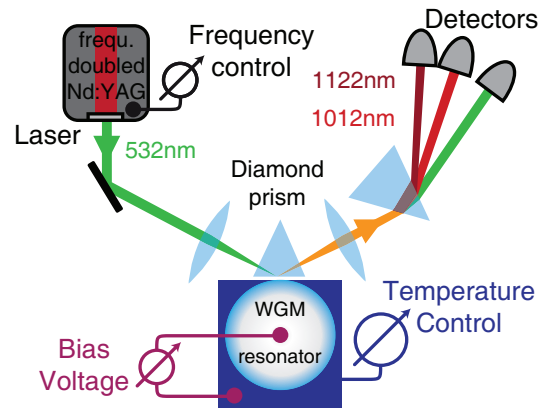}
	\caption{(Color online.) Experimental setup for PDC. }
	\label{fig:setup}\vspace*{-0.2in}
\end{figure}

For investigating these intriguing properties of a WGM OPO we built a setup schematically shown in Fig.~\ref{fig:setup}. This setup is similar to the one used in \cite{furst10}. It consists of a diamond turned and polished disk resonator made from a congruent $5$\% MgO-doped $z$-cut Lithium Niobate wafer of $0.5$ mm height. The disk radius is $R=1.9$
mm and the rim radius is 0.25 mm. 

A frequency-doubled Nd:YAG laser beam is coupled evanescently to our resonator via a diamond prism using frustrated total internal reflection. Adjusting the distance between the prism and the resonator we change the overlap between their evanescent fields. This allows us to tune the coupling rate $\gamma_0$ of the WGM resonator, which would be equivalent to realizing mirrors with continuously variable reflectivity in usual optical resonators.

The optical axis of the crystal is chosen to coincide with the symmetry axis of the resonator. In this geometry noncritical Type I PDC phase matching can be achieved between the
extraordinary polarized pump ($p$) with a wavelength of 532 nm and ordinary polarized signal ($s$) and idler ($i$) with wavelengths near 1064 nm at a crystal temperature near $94^{\circ}$C. The triple resonance of the WGM resonator with the pump field and the two parametric fields is achieved by using the temperature dispersion and the electro-optical effect, as discussed in \cite{furst10}. Our measurement does not require any locking mechanisms, as the monolithic WGM resonators are remarkably stable. 
The non-degenerate signal and idler beams are coupled out by the same coupling prism, and separated for detection by a dispersion prism.

\begin{figure}[t]
\centering
	\includegraphics[width=7cm]{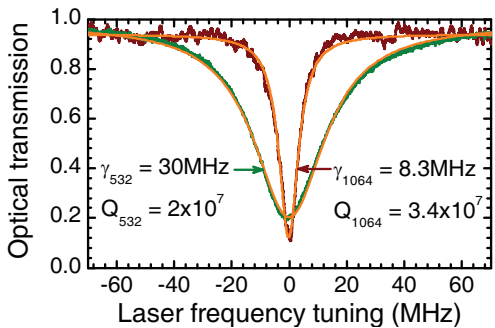}
	\caption{(Color online.) WGM resonances at 532 nm and 1064 nm wavelengths. 
}
	\label{fig:2}\vspace*{-0.1in}
\end{figure}

To study the quality factor of the resonator, we measured the critically coupled WGMs linewidths $\gamma$ at the pump and parametric wavelengths.  
In Fig.~\ref{fig:2} we show two such modes with $Q_{s,i}\approx 3.4\times 10^7$ at 1064 nm, and $Q_{p}\approx 2.0\times 10^7$ at 532 nm.
The measurement of $Q$-factors reveals the resonator's intrinsic loss nature. The two dominant loss mechanisms in our resonator are material absorption and Rayleigh scattering at the resonator's defects and surface roughness. Radiation losses for this size of the resonator are negligible. Since the Rayleigh loss scales as the fourth power of the wavelength, one would expect a 16-fold increase of $Q$ for twice the wavelength. Instead, we see an approximately 1.7-fold augmentation, attributed to stronger absorption of LiNbO$_3$ for shorter wavelengths \cite{kuz}. Thus, our resonator quality factor is limited by material absorption only.

The critital coupling contrast of 80\% is limited by matching of the pump footprint spatial profile to the WGMs evanescent field in the coupling region. In the following we will operate with the ``in-coupled" pump power, which is equal to 80\% of the incident pump power.

\begin{figure}[t]
\centering
\includegraphics[width=7.5cm]{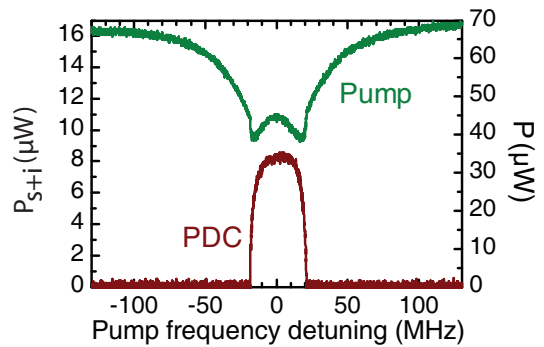}
	\caption{(Color online.) PDC and pump output vs. pump frequency detuning. 
}
	\label{fig:pdc}\vspace*{-0.1in}
\end{figure}

\begin{figure}[b]
\centering
\includegraphics[width=7cm]{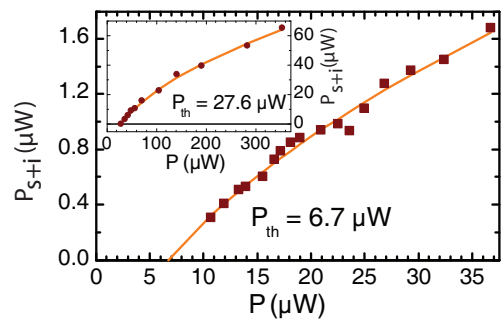}
	\caption{(Color online.) The measurement and theoretical fit for the resonant OPO output vs. in-coupled pump for a critically (inset) and weakly coupled resonator. 
}
	\label{fig:p}\vspace*{-0.1in}
\end{figure}

In contrast with Fig.~\ref{fig:2}, the pump WGM in Fig.~\ref{fig:pdc} is phase-matched to the parametric WGMs. As soon as the in-coupled power exceeds the threshold value, the PDC fields are generated, which in turn affects the pump intensity. A series of such measurements for the same mode is summarized in Fig.~\ref{fig:p}, which shows the total OPO output power $P_{s+i}$ at the exact resonance as a function of the in-coupled pump power $P$. Note that the conversion efficiency is larger in the stronger coupled case.

Following the theoretical analysis of the triply-resonant OPO dynamics \cite{variance}, we derive the following relation:
\begin{equation}
P_{s+i}=8P_{th}\frac{\gamma_{p0}}{\gamma_{p}}\frac{\gamma_0}{\gamma}
\left(\sqrt{P/P_{th}}-1\right),
\label{output}
\end{equation} 
where $\gamma_{p0}/\gamma_{p}$ is the ratio of the pump WGM coupling rate to its linewidth, and $\gamma_{0}/\gamma$ is a similar ratio for the signal and idler. Fitting the data in Fig.~\ref{fig:p} with Eq.~(\ref{output}) we find $P_{th}$ for an undercoupled and a critically coupled pump. The lowest threshold power was estimated to be 6.7 $\mu$W. While this is a significant improvement over previously reported values of 300 $\mu$W for PDC \cite{threshold} and 170 $\mu$W for a four-wave-mixing process \cite{kippenberg04kerr}, the measured threshold is still much higher than expected. Partly this may be due to the discrepancy between the spherical approximation adopted in our analysis and non-spherical geometry of our resonator. However, the main reason appears to be the non-optimal mode overlap, which can occur in many ways (see Fig.~\ref{fig:radial}).  
Therefore finding the optimal parametrically coupled WGMs should again increase the OPO efficiency by more than two orders of magnitude. 


\begin{figure}[t]
\centering
\includegraphics[width=8cm]{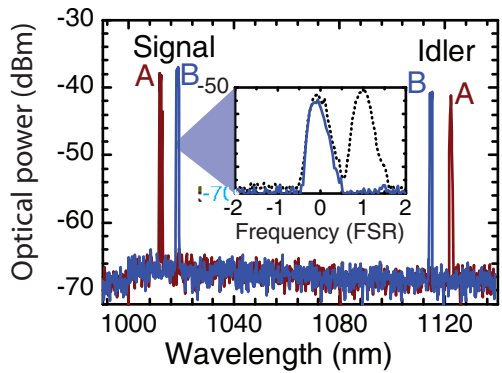}
	\caption{(Color online.) Optical spectrum of the parametric light: two different mode pairs (A and B) are excited under different conditions. Inset: a zoom on one spectral component proves single-mode operation (solid and dashed lines represent diffirent conditions).
}
	\label{fig:spectrum}\vspace*{-0.1in}
\end{figure}

For many spectroscopy and quantum optics applications, narrow-line (ideally, single frequency mode) OPOs are desirable. 
We recorded the emission spectra of our WGM OPO for different phase matching conditions using an optical spectrum analyzer. Two such spectra labeled A and B are shown in Fig.~\ref{fig:spectrum}. As seen from the figure, the signal and idler are separated by more than 100 nm. Moreover, each parametric beam consists of a single frequency mode. To verify it, we 
recorded the signal and idler spectra with a resolution of approximately 2.7 GHz, which is sufficient to distinguish two WGMs separated by the signal and idler free spectral range (FSR) of approximately 12 GHz. While recording this spectrum, the pump frequency sweeps across approximately 500 MHz, and the optical spectrum analyzer captures and retains all signal frequencies generated during the sweep.  The result is shown in the inset of Fig.~\ref{fig:spectrum}. The solid trace recorded over several minutes with no active locking of the resonator shows only one mode. For comparison, we have forced the resonator to ``mode hop" across one FSR while recording the dashed trace, and verified that the two-mode operation would be clearly visible on the spectrum analyzer. Thus we conclude that our OPO operates in a single mode regime.

The single-mode operation is a consequence of the FSR dispersion $\Delta(\lambda)=c/ 2\pi Rn(\lambda)$ which arises from the refractive index dispersion $n(\lambda)$. If the FSR dispersion is smaller than the signal and idler cavity bandwidth, i.e. if $\vert \Delta_i-\Delta_s \vert <\gamma$,  multi-mode PDC is allowed by the phase matching and energy conservation conditions. Then an FSR-spaced comb of pair-wise matched signal and idler frequencies is expected, centered at the degenerate PDC frequency. 
However, in a highly nondegenerate PDC the FSR dispersion is strong, $\vert \Delta_i-\Delta_s \vert >\gamma$, and the single-mode regime is enforced by the energy conservation.

The difference of the signal ($\lambda_s\approx 1010$ nm) and idler ($\lambda_i\approx 1120$ nm) FSRs in our resonator can be found from the Sellmeier equation \cite{Schlarb94}: $\Delta_i-\Delta_s\approx 0.2$ GHz, or from a direct measurement using the optical spectrum analyzer: $\Delta_i-\Delta_s = 0.16\pm 0.09$ GHz. The theoretical and experimental results agree, and both greatly exceed even the overcoupled WGM's linewidth. Hence, the single-mode operation of our OPO is to be expected. 
 
To summarize, we have demonstrated a strongly non-degenerate, single mode, narrow line, high stability OPO with an extremely low threshold, based on a second-order nonlinear WGM resonator. Our analysis reveals the role of the photon orbital momentum conservation in the PDC phase matching in spherical geometry, explained the presence of multiple PDC channels with different efficiencies, and predicted that even lower OPO thresholds could be achieved.
This work illustrates the great potential of WGM-based OPOs. 
They will allow for investigating the rich dynamic phenomena in far above-threshold regime, and will provide the underlying technology for compact, stable, low-power tunable spectroscopy sources with narrow linewidth. Utilizing quasi phase matching in periodically poled resonators \cite{ilchenko04SH,sasagawa} one could achieve flexibility in the selection of wavelengths. Finally, these resonators open the possibility to combine the optical nonlinearity with optomechanics, where the crystalline materials are a promising candidate \cite{kippmech}.

The authors would like to acknowledge funding from EU project COMPAS. D.V.S. acknowledges funding from the Alexander von Humboldt foundation, and J.U.F. from IMPRS. J.U.F. and Ch. M. thank Alessandro Villar for helpful discussions. 



\begin{thebibliography}{99}

\bibitem{vahala1}
K. J. Vahala, 
Nature \textbf{424}, 839 (2003).

\bibitem{ilchenko_app}
V. S. Ilchenko, A. B. Matsko, 
IEEE J. Quant. Elect. \textbf{12}, 15 (2006).

\bibitem{optomech}
G. Anetsberger \textit{et al.}, 
Nat. Photonics \textbf{2}, 627 (2008);
F. Marquardt \textit{et al.}, 
Physics \textbf{2}, 40 (2009).

\bibitem{gotzinger}
S. G\"otzinger \textit{et al.}, 
Nano Lett. \textbf{6}, 1151 (2006).

\bibitem{aoki}
T. Aoki \textit{et al.}, 
Nature \textbf{443}, 671 (2006).

\bibitem{vahala_rev}
T. J. Kippenberg and K. J. Vahala, 
Science \textbf{321}, 1172 (2008).

\bibitem{grudinin}
I. S. Grudinin \textit{et al.}
Phys. Rev. Lett. \textbf{104}, 083901 (2010).
    
\bibitem{kippenberg_4wave} T. J. Kippenberg \textit{et al.}
Phys. Rev. Lett. \textbf{93}, 083904 (2004).

\bibitem{carmon05_locking}
T. Carmon \textit{et al.}, Opt. Express, \textbf{13}, 3558-3566 (2005).
	 
\bibitem{delhaye_4wave} P. Del'Haye \textit{et al.}
Nature \textbf{450}, 1214 (2007).

\bibitem{savchenkov_4wave} A. A. Savchenkov \textit{et al.}, 
Phys. Rev. Lett. \textbf{93}, 243905 (2004).

\bibitem{savchenkov_07}	
A. A. Savchenkov \textit{et al.},
Opt. Lett. \textbf{32}, 157 (2007). 

\bibitem{ilchenko04SH}
 V. S. Ilchenko \textit{et al.}, 
 Phys. Rev. Lett. \textbf{92}, 043903 (2004). 
 
\bibitem{furst10}
J. U. F\"urst \textit{et al.}, 
Phys. Rev. Lett. \textbf{104}, 153901 (2010).

\bibitem{ilchenko03parametric}
V. S. Ilchenko \textit{et~al.} 
J. Opt. Soc. Am. B, \textbf{20}, 1304 
(2003).

\bibitem{Kozyreff08}
G. Kozyreff \textit{et al.}, 
Phys. Rev. A \textbf{77}, 043817 (2008). 
 
\bibitem{Gaunt29}
J.~A. Gaunt,
 Phil. Trans. R. Soc. A \textbf{228}, 151 (1929).
  
\bibitem{kuz}
Y. S. Kuz'minov, \textit{Lithium Niobate Crystals}, (Cambridge International Science Publishing 1999).

\bibitem{variance}
C. Fabre \textit{et al.}, 
J. Phys. France \textbf{50}, 1209 (1989).

\bibitem{threshold}
K. S. Zhang \textit{et al.}, 
Phys. Rev. A \textbf{64}, 033815 (2001).

\bibitem{kippenberg04kerr}
T. J. Kippenberg \textit{et al.}, 
Phys. Rev. Lett. \textbf{93}, 083904 (2004).

\bibitem{Schlarb94} 
U. Schlarb and K.Betzler, 
Phys. Rev. B \textbf{50}, 751 (1994).

\bibitem{sasagawa}
K. Sasagawa and M. Tsuchiya, 
Conference on Lasers and Electro-Optics 
(CLEO) paper CThC6 (2007).

\bibitem{kippmech}
J. Hofer \textit{et al.}, 
arXiv:0911.1178v2 (2009).


\end{thebibliography}
\end{document}